\documentstyle[aps,prl,preprint,floats,epsfig]{revtex}

\begin{document}

\tighten

\preprint{\tighten\vbox{\hbox{\bf CLNS 02-1787}
                        \hbox{\bf CLEO 02-7}}}

\title{\Large \boldmath Correlated
inclusive $\Lambda{\overline\Lambda}$ production
in $e^+e^-$ annihilations at $\sqrt{s}\sim$10.5 GeV}

\author{CLEO Collaboration}
\date{\today}

\maketitle
\tighten

\begin{abstract}
Using a 
13.7 ${\rm fb}^{-1}$ sample
of continuum two-jet
$e^+e^-\to q{\overline q}$ events
collected with the CLEO detector, we have searched for
correlations between $\Lambda$ and ${\overline\Lambda}$ particles,
specifically in cases where the opening angle between the two 
particles is large and each has momentum $>$1 GeV/c.
Such correlations may
indicate the presence of
baryon number conservation at the primary quark level.
A previous CLEO study of $\Lambda_c{\overline\Lambda_c}$
correlations\cite{lclcbcorrelations} 
indicated direct, associated production of 
primary charmed 
baryons $\Lambda_c$:
$e^+e^-\to c\overline{c}\to\Lambda_c{\overline\Lambda_c}$.
That effect was not observed in Monte Carlo simulations.
Our current search for similar direct,
associated production of $\Lambda$ baryons at the primary quark
level ($e^+e^-\to s\overline{s}\to\Lambda{\overline\Lambda}$, e.g.)
qualitatively indicates a similar 
effect, although it relies on a Monte Carlo dependent
subtraction of background 
$\Lambda{\overline\Lambda}$ production (based on the default
JETSET 7.4 event generator).
\end{abstract}

\pacs{\em 13.30.-a, 13.60.Rj, 13.65.+i, 14.20.Lq}
\setcounter{footnote}{0}

\newpage

\begin{center}
Z.~Metreveli,$^{1}$ K.K.~Seth,$^{1}$ A.~Tomaradze,$^{1}$
P.~Zweber,$^{1}$
S.~Ahmed,$^{2}$ M.~S.~Alam,$^{2}$ L.~Jian,$^{2}$ M.~Saleem,$^{2}$
F.~Wappler,$^{2}$
E.~Eckhart,$^{3}$
% ...
\end{center}
 
\small
\begin{center}
$^{1}${Northwestern University, Evanston, Illinois 60208}\\
$^{2}${State University of New York at Albany, Albany, New York 12222}\\
$^{3}${Ohio State University, Columbus, Ohio 43210}\\
% ...
\end{center}


\section{Introduction}
At $\sqrt{s}\sim$10 GeV, below the threshold for
$e^+e^-\to B{\overline B}$,
particle production in $e^+e^-$ annihilation occurs in a largely
low-$Q^2$, non-perturbative 
regime. Fragmentation models are therefore
used to describe the process
$e^+e^-\to$hadrons. Computer codes such as JETSET\cite{lund92} have been
extremely successful at matching 
experimental results on
inclusive particle production, multiplicities
and angular distributions, both qualitatively and, to a 
large degree, quantitatively.
Compensation of baryon number is one of the more
subtle aspects of particle
fragmentation modeling.
One obvious question is whether baryon compensation
occurs locally (e.g., small opening angle between baryon
and antibaryon)
or globally (large opening angles).
In the case when a baryon is
produced in the first step of fragmentation ($e^+e^-\to c{\overline c}$;
$c\to\Lambda_c$, e.g.), it is possible that both baryon and flavor quantum
numbers will be compensated in the opposite hemisphere\footnote{We
define two particles to be  
``opposite hemisphere'' if their opening angle exceeds 90 degrees.
This definition is therefore disconnected from momentum flow in the
remainder of the event and is only indirectly related to 
such standard parameters as thrust, sphericity, etc. As is demonstrated
in the text, most $\Lambda$
baryons emerge either very close to, or directly opposite,
the anti-$\Lambda$ studied, so the separation of an event into
hemispheres is at least approximately valid.}
(e.g., $e^+e^-\to c{\overline c}$;
$c\to\Lambda_c$; ${\overline c}\to{\overline\Lambda_c}$). 
However, in 
models in which the primary quark and anti-quark
fragment entirely independently of each other, no 
such correlations are
expected. 

Previous LEP studies \cite{DELPHI-paper,ALEPH-paper,OPAL-paper}
of $\Lambda{\overline\Lambda}$ production
at $\sqrt{s}$=90 GeV found that in events containing both a
$\Lambda$ and a ${\overline\Lambda}$, the two particles tend to be
produced with small opening angles between them.
 Those analyses also discrimininated between different models of
 $\Lambda{\overline\Lambda}$ production. It was found
in the DELPHI and ALEPH analyses that the JETSET
 string fragmentation event generator gave excellent agreement with
 data when the ``popcorn'' control parameter $\rho$ was set to 0.5; the
OPAL analysis favored slightly higher values of $\rho$. (In
the default CLEO
Monte Carlo event generator, we have used $\rho$=0.5.)
In those previous studies, no statistically
significant signal for correlated opposite-hemisphere, primary
($\Lambda{\overline\Lambda}$) production was found.

A previous CLEO study\cite{lclcbcorrelations} of charmed baryons sought
to discriminate between independent vs. correlated fragmentation models. 
Assuming that primary particles fragment independently, 
then the number of times that we find 
a $\Lambda_c$ baryon opposite a ${\overline\Lambda}_c$ 
antibaryon in an event (denoted ``$\Lambda_c|{\overline\Lambda_c}$''), 
scaled to the total number of observed
${\overline\Lambda_c}$ 
(denoted ``$\frac{({\Lambda}_{c} | {\overline\Lambda_c})}
{{\overline\Lambda_c}}$''), 
should be equal to the number of times that we find a 
${\Lambda_c}$ baryon opposite any other anti-charmed 
hadron ${\overline H_c}$ 
scaled to the total number of observed anti-charmed hadrons
($\frac{({\Lambda}_{c} 
| \overline{H_{c}})}{\overline{H_{c}}}$).
It was found that,
 given a ${\overline\Lambda_c}$ 
(reconstructed in five different decay modes),
a $\Lambda_c$ is observed in the opposite hemisphere 
$(0.72\pm0.11)$\% of the time (not corrected for efficiency).
By contrast, given a ${\overline D}$,
a $\Lambda_c$ is observed in the opposite hemisphere only 
$(0.21\pm0.02)$\% of the
time. Normalized to the total number of either ${\overline\Lambda_c}$ or
${\overline D}$ ``tags'', that study concluded that
it is 3.52$\pm$0.45$\pm$0.42 
times more likely to find a $\Lambda_c$ opposite
a ${\overline\Lambda_c}$ than opposite a ${\overline D}$ meson.
This enhancement is not produced in the
default JETSET 7.4 $e^+e^-\to c{\overline c}$
Monte Carlo simulation.

As a straightforward extension of that analysis, we can search for
similar correlations between 
primary $\Lambda$ and ${\overline\Lambda}$ baryons.
In this case, the correlation is obscured by the fact that, unlike
$\Lambda_c$ baryons, $\Lambda$'s produced in $e^+e^-$ annihilations
do not necessarily contain the primary quarks, and will be produced
copiously in fragmentation, as well as in weak decays of charmed baryons. 
In our current analysis,
the production of 
($\Lambda|{\overline\Lambda}$) through fragmentation is modeled using the
JETSET 7.4 event generator combined
with a GEANT-based\cite{GEANT} 
simulation of our detector.
The ``feeddown'' contribution from correlated
primary ($\Lambda_c|{\overline\Lambda_c}$) production
($\Lambda_c\to\Lambda |{\overline\Lambda_c}\to{\overline\Lambda} $)
is evaluated from the data itself.

For this study, 
$\Lambda_c^+$'s are fully
reconstructed in the decay modes
$\Lambda_c^+\to pK^-\pi^+$, $\Lambda_c^+\to pK^0_S$, 
$\Lambda_c^+\to\Lambda\pi^+$, $\Lambda_c^+\to\Lambda\pi^+\pi^-\pi^+$,
$\Lambda_c^+\to pK^0_S\pi^+\pi^-$,\footnote{Charge conjugation is implicit.}
and partially reconstructed
through $\Lambda_c^+\to\Lambda$X.
$\Lambda$ baryons are reconstructed in $\Lambda\to p\pi^-$.

\section{Apparatus and Event Selection}
\label{sec:event_selection}

This analysis was performed using the CLEO II and the upgraded
CLEO II.V detectors operating at the
Cornell Electron Storage Ring (CESR) at center-of-mass energies $\sqrt{s}$
= 10.52--10.58 GeV.  
The event sample 
used for this measurement is comprised of 9.2 ${\rm fb}^{-1}$ of data
collected at the $\Upsilon$(4S) resonance and 4.6 ${\rm fb}^{-1}$ of data 
collected about 60 MeV below the $\Upsilon$(4S) resonance. Approximately
$20\times 10^6$ continuum
$c{\overline c}$ events\footnote{Corresponding to approximately
$20\times 10^6$ $u{\overline u}$, $5\times 10^6$ $d{\overline d}$, and
$5\times 10^6$ $s{\overline s}$ events,
respectively.} 
are included in this sample.

For 4.6 fb$^{-1}$ of the data used for this analysis 
(``CLEO-II'' data\cite{kubota92}),
measurements of charged particle momenta were made with
three nested coaxial drift chambers consisting of 6, 10, and 51 layers,
respectively. In a subsequent upgrade (``CLEO-II.V''\cite{CLEO-IIV},
corresponding to the remaining data
used for this analysis), 
the innermost tracking chamber was replaced with a high-precision
silicon detector, and the gas in the main tracking volume was changed
to provide better cell resolution and improved specific ionization
(dE/dx) resolution\cite{DR-gas}.
The entire tracking system fills the volume from $r$=3 cm to $r$=1 m, with
$r$ the radial coordinate relative to the beam (${\hat z}$) axis. 
This system is very efficient ($\epsilon\ge$98\%) 
for detecting tracks that have transverse momenta ($p_T$)
relative to the
beam axis greater than 200 MeV/c, and that are contained within the good
fiducial volume of the drift chamber ($|\cos\theta_Z|<$0.94, with $\theta_Z$
defined as the polar angle relative to the beam axis). 
This system achieves a momentum resolution of $(\delta p/p)^2 =
(0.0015p)^2 + (0.005)^2$ ($p$ is the momentum, measured in GeV/c). 
Pulse-height measurements in the main drift chamber provide specific
ionization resolution
of 5.0\% (CLEO II.V) or 5.5\% (CLEO II) 
for Bhabha events, giving excellent 
$K/\pi$ separation for tracks with
momenta up to 700 MeV/c and separation of order 2$\sigma$ in the relativistic
rise region above 2.5 GeV/c. 
Outside the central tracking chambers are plastic
scintillation counters, which are used as a fast element in the trigger system
and also provide particle identification information from 
time-of-flight measurements.  

Beyond the time-of-flight system is the electromagnetic calorimeter,
consisting of 7800 thallium-doped CsI crystals.  The central ``barrel'' region
of the calorimeter covers about 75\% of the solid angle and has an energy
resolution which is empirically found to follow:
\begin{equation}
\frac{ \sigma_{\rm E}}{E}(\%) = \frac{0.35}{E^{0.75}} + 1.9 - 0.1E;
                                \label{eq:resolution1}
\end{equation}
$E$ is the shower energy in GeV. This parameterization includes
noise effects, and translates to an
energy resolution of about 4\% at 100 MeV and 1.2\% at 5 GeV. Two end-cap
regions of the crystal calorimeter extend solid angle coverage to about 95\%
of $4\pi$, although energy resolution is not as good as that of the
barrel region. 
The tracking system, time-of-flight counters, and calorimeter
are all contained 
within a superconducting coil operated at 1.5 Tesla. 
An iron flux return interspersed with
proportional tubes
used for muon detection are located immediately outside the coil and 
in the two end-cap regions.

Primary
proton, kaon or pion charged track candidates 
must pass the following restrictions:

(a) The track 
must pass a 99\% probability criterion for its assumed
particle identification,
based on the associated 
charged track's specific ionization measured in the drift chamber.

(b) The track must have momentum greater than 100 MeV/c.

Each reconstructed charmed hadron must have momentum greater than
2.3 GeV/c to ensure that there is no contamination from $B$-meson decays
to charm.

\section{Data Analysis}
\subsection{Production Ratios}

We define the single-tag yield to be the 
number of reconstructed events containing one 
particular hadron $H$.
This yield 
is typically determined by
fitting a double-Gaussian signal atop a smooth, 
low-order polynomial background
function.
The number of double-tags is 
defined as the number of 
events in which two specific particles are both reconstructed,
separated by less than 90 degrees ($H_1{\overline H_2}|$; `same-hemisphere') or
greater than 90 degrees
($H_1|{\overline H_2}$; `opposite hemisphere').\footnote{We use the notation
``$H_1H_2$'' (without a vertical bar) to indicate a generic
correlation event in which 
the two particles may be found in either hemisphere relative to each other.}

To suppress possible contributions from 
$e^+e^-\to\Upsilon(4S)\to B{\overline B}$, with 
subsequent decays such as
${\overline B}\to\Xi_c(\to\Lambda X){\overline\Lambda} X$, we have imposed
a minimum momentum requirement $p_{\Lambda}>$1.0 GeV/c. Decays of 
$B$-mesons should generally produce lower momentum $\Lambda$ 
and ${\overline\Lambda}$; 
$B{\overline B}$ events are also likely to have different 
$\Lambda{\overline\Lambda}$ production dynamics compared to
$\Lambda{\overline\Lambda}$ production resulting from direct
$e^+e^-$ annihilations.
(The requirement $p_{\Lambda}>$1.0 GeV/c is ``standard'' in other
continuum Lambda studies as 
well\cite{lclcbcorrelations,lamctolamx,pkpipaper}.)
In order to check whether there is $B{\overline B}$ contamination of 
the $\Lambda{\overline\Lambda}$ data sample, we compare our
$\Lambda{\overline\Lambda}$ yield derived from data taken on the 
$\Upsilon$(4S) resonance with
the  $\Lambda{\overline\Lambda}$ yield obtained using data taken on the
continuum below the resonance. We expect these yields to be in the
ratio of the integrated luminosities of the two samples, corrected for the 1/s
dependence of the continuum cross-section, if there is no 
$B{\overline B}$
contamination (these two effects give an expected ratio of 2.01 for
our data).
Requiring $p_{\Lambda}>$1.0 GeV/c, the yield of
same-hemisphere $\Lambda{\overline\Lambda}|$ pairs on the $\Upsilon$(4S)
resonance ($5610.4\pm79.1$) compared to the yield for the 
continuum
($2815.2\pm56.1$) is, indeed, consistent (at the $2\sigma$ level) 
with this ratio.
Opposite hemisphere yields ($4962.8\pm78.8$ and $2398.9\pm54.4$,
respectively) are similarly consistent with an exclusively continuum
origin of our  $\Lambda|{\overline\Lambda}$ sample.
For maximal statistics, we use our entire sample for the subsequent
analysis, and discuss residual $B{\overline B}$ contamination effects
later in this document.

\begin{figure}[htpb]
\begin{picture}(200,250)
\includegraphics{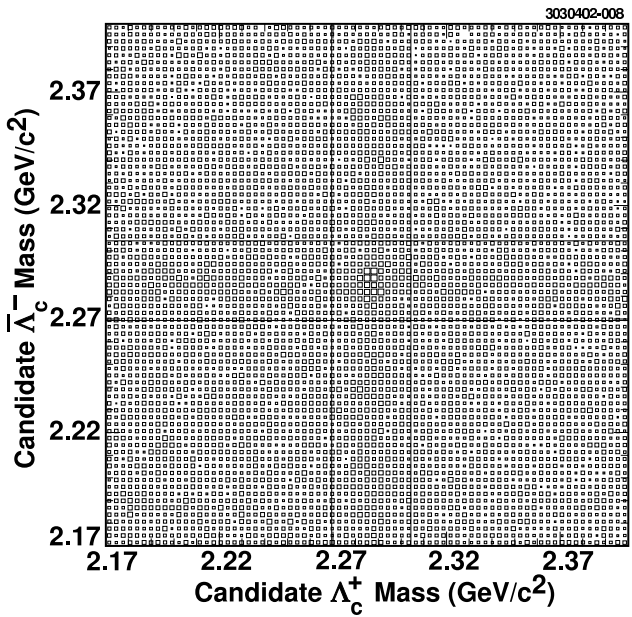}
\end{picture} \\
\caption{\small 
Double-tag invariant mass
of $\Lambda_c$ candidates
plotted vs. invariant mass of ${\overline\Lambda_c}$ candidates 
(${\Lambda_c} | {\overline\Lambda_c}$) from data. Shown is the
sum of the modes: 
$\Lambda_c^+\to pK^-\pi^+$, $\Lambda_c^+\to pK^0_S$, 
$\Lambda_c^+\to\Lambda\pi^+$, $\Lambda_c^+\to\Lambda\pi^+\pi^-\pi^+$,
and $\Lambda_c^+\to pK^0_S\pi^+\pi^-$ (and their charge conjugates,
in the case of ${\overline\Lambda_c}$ reconstruction).}
\label{fig:lclcb}
\end{figure}

\begin{figure}[htpb]
\begin{picture}(200,250)
\includegraphics{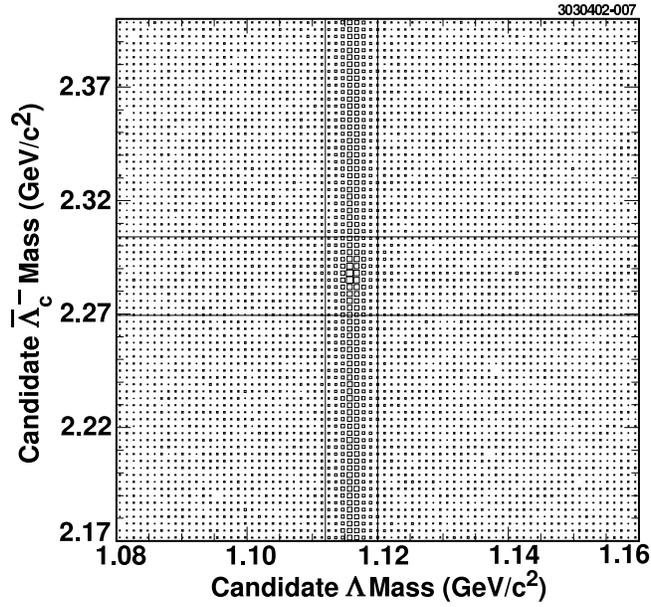}
\end{picture} \\
\caption{\small Double-tag plot of 
$\Lambda | {\overline\Lambda_c}$ (plus charge conjugate) from data.  
The ${\overline\Lambda_c}$ is selected as in the previous Figure; the
$\Lambda$ is reconstructed in $\Lambda\to p\pi^-$. }
\label{fig:lcl0b}
\end{figure}

\begin{figure}[htpb]
\begin{picture}(200,250)
\includegraphics{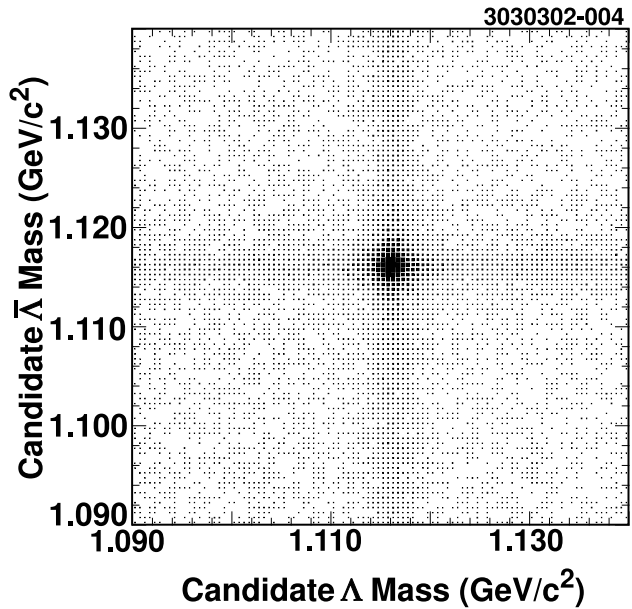}
\end{picture} \\
\caption{\small Double-tag plot of $\Lambda | \overline{\Lambda}$
(opposite hemisphere) for data.}
\label{fig:l0l0bback}
\end{figure}

\begin{figure}[htpb]
\begin{picture}(200,250)
\includegraphics{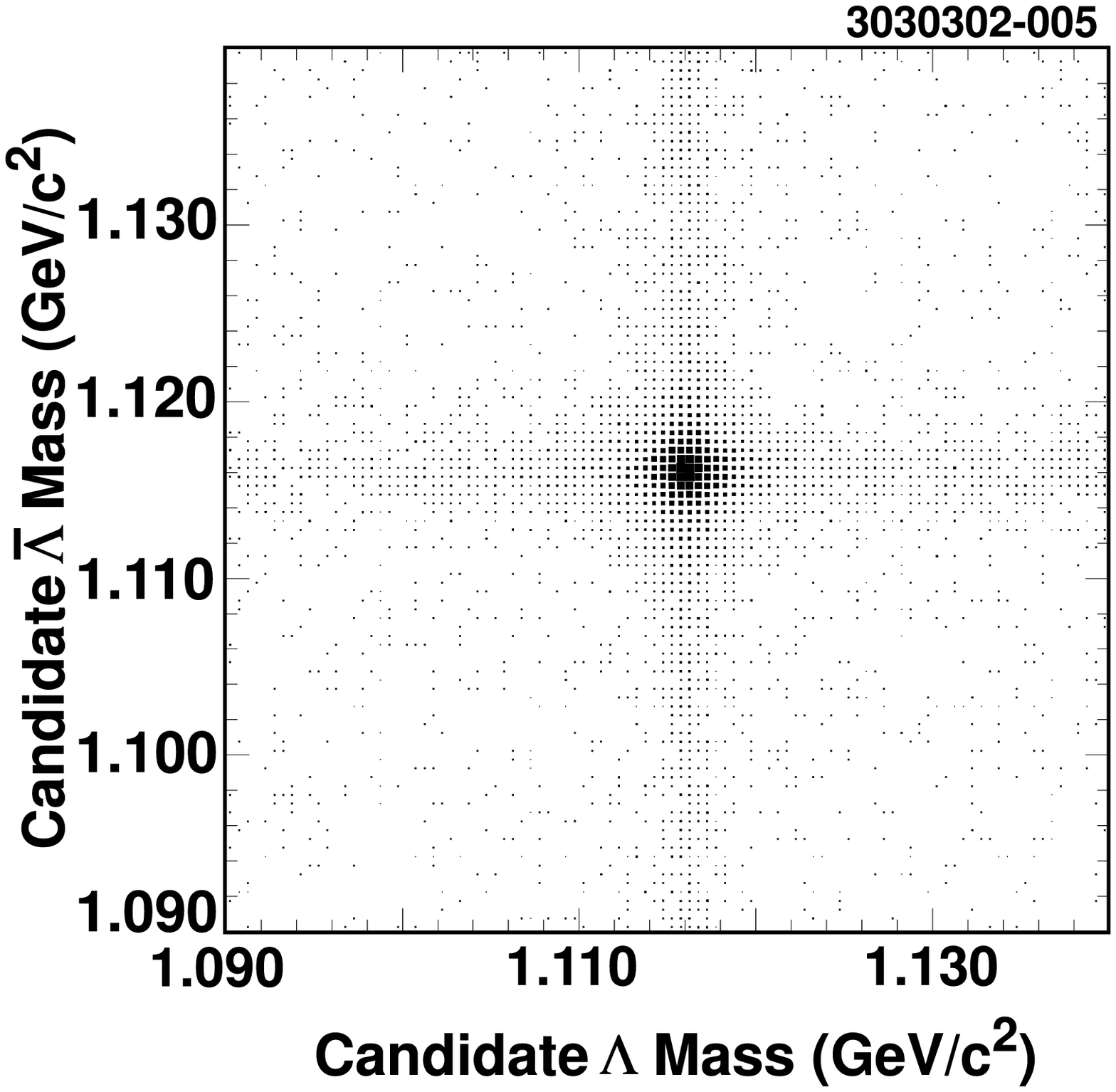}
\end{picture} \\
\caption{\small Double-tag plot of 
${\Lambda} \overline{\Lambda}|$ (same hemisphere) for data.}
\label{fig:l0l0bfor}
\end{figure}

The yields of 
$\Lambda_c|{\overline\Lambda_c}$, $\Lambda_c|{\overline\Lambda}$,
$\Lambda|{\overline\Lambda}$ (opposite hemisphere) and 
$\Lambda{\overline\Lambda}|$ (same hemisphere)
double-tags
are extracted from two-dimensional 
invariant mass plots, shown in Figures \ref{fig:lclcb},
\ref{fig:lcl0b}, and \ref{fig:l0l0bback}, and
\ref{fig:l0l0bfor}, 
respectively.
The total correlated double-tag yield is 
first determined by 
fitting one-dimensional projections of the two-dimensional plots. 
Consider, for example Fig. \ref{fig:lclcb}.
We take one-dimensional projections of three slices in the
candidate ${\overline\Lambda_c}$ recoil
invariant mass:
the ${\overline\Lambda_c}$ signal region:
($|M_{recoil}-2.286|<$0.03 $GeV/c^2$)
and the two ${\overline\Lambda_c}$ sideband regions:
($0.04<|M_{recoil}-2.286|<$0.07 $GeV/c^2$).
We then subtract the fitted $\Lambda_c$ yields
from the ${\overline\Lambda_c}$ 
sidebands from that of the signal region.
In performing these fits, the double-Gaussian signal shapes are 
constrained using the parameters determined from fits to the
single-tag sample. 
We also perform a single fit in two dimensions to extract the signal yields.
In this latter fit, a two-dimensional Gaussian signal is used to parametrize
the peak region, two single Gaussians are used to fit the ``ridges'' away
from the peak region (corresponding to true signals along one
axis in association
with combinatoric background on the other axis)
and a two-dimensional, smooth polynomial is used to
parametrize the background. The two procedures result in consistent
signal yields; the yields
presented in Table I result from application of the second procedure.

Table \ref{tab:lamyields} summarizes
yields in both data and Monte Carlo simulations of comparable
size. No detection efficiency
corrections have been applied. The number of observed
inclusive, single-tag particles is presented, in addition to the
number of same hemisphere double-tags 
(`$\Lambda{\overline\Lambda}|$'), opposite-hemisphere double-tags
(`$\Lambda|{\overline\Lambda}$'), and the rate of 
double-tags per single-tag (`$(\Lambda|{\overline\Lambda})/\Lambda$'). 
Where appropriate, differences of data ratios minus Monte Carlo ratios 
are given to allow direct comparisons (column 4 in the Table).

\message{CAN WE TAG THE PRIMARY MC LAMBDA-LAMBDABAR YIELD?}
\begin{table}
\caption{\label{tab:lamyields}Yields in data vs. Monte Carlo simulations.
The data are drawn from a sample of $\sim 50\times 10^6$ 
hadronic events ($\sim$2/3
on-$\Upsilon$(4S) events plus $\sim$1/3 taken on the continuum
below the $\Upsilon$(4S)); the MC is
drawn from a sample of $\sim 60\times 10^6$ 
exclusively continuum events generated
by JETSET 7.4, fully simulated in our CLEO detector and reconstructed
using the same algorithms as applied to data. 
Derivation
of final results (specifically, the derivation of 
the $\Lambda|{\overline\Lambda}$ (q\=q) signal yield and evaluation of the
non-primary $\Lambda|{\overline\Lambda}$ (np) and 
charmed-baryon $\Lambda|{\overline\Lambda}$ (c\=c) backgrounds)
are discussed in detail in the text.
The difference (Data-Monte Carlo) in the fourth column is shown only for
the yield ratios, which are sample-size independent.} 
\begin{center}
\begin{small}
\begin{tabular}{lcc||r} & Data & Monte Carlo & Data - MC \\ \hline
                                                         
 $\Lambda_c+{\overline\Lambda_c}$ ($p>$2.3 GeV/c) & 83955 $\pm$ 1852 & 
92731 $\pm$ 810 & \\ \hline

($\Lambda+{\overline\Lambda}$) (ON-4S, $p>$ 1 GeV/c)  & $491087\pm822$ & \\
($\Lambda+{\overline\Lambda}$) 
(Continuum, $p>$ 1 GeV/c) & $236162\pm575$ & $973657\pm1121$ & \\ \hline

$\Lambda+{\overline\Lambda}$ (total, $p>$ 1 GeV/c) 
& $727249\pm1003$ & $973657\pm1121$ & \\ \hline \hline

($\Lambda_c | {\overline\Lambda_c}$) & 323.6$\pm$39.3 &  97.2 $\pm$ 40.4 & \\
($\Lambda_c|{\overline \Lambda}$) +  (${\overline\Lambda_c}|{\Lambda}$)
& $1470.6\pm74.1$ & $695.8\pm67.1$ & \\

 ($\Lambda_c {\overline \Lambda}|$) +
 (${\overline\Lambda_c}{\Lambda}|$) & $249.4\pm26.2$ & $210.7\pm28.9$ & \\

(${\Lambda} | {\overline \Lambda}$) & $7361.8\pm95.7$ & $6519\pm91.0$ \\

($\Lambda {\overline \Lambda}|$)  & $8425.6\pm97.0$   & $11394.9\pm112.4$ & \\ 

($\Lambda {\Lambda}|$)  & $28.6\pm15.3$ & $36.4\pm12.4$ & \\             

($\Lambda | {\Lambda}$)  & $239.8\pm37.2$ & $401.0\pm33.9$  & \\ \hline \hline

($\Lambda_c | {\overline\Lambda_c}$)/($\Lambda_c+{\overline\Lambda_c}$) 
($\times 10^{-3}$) &  $3.91\pm0.47$  &  $1.05\pm0.44$ & $2.86\pm0.64$ \\

(${\overline\Lambda_c}{\Lambda}|+\Lambda_c{\overline\Lambda}|)/(\Lambda+{\overline\Lambda})$ 
$(\times 10^{-4})$ & 
$3.43\pm0.36$ & $2.16\pm0.30$ &  $1.27\pm0.46$ \\ 

(${\overline\Lambda_c}|{\Lambda} + \Lambda_c|{\overline\Lambda})/(\Lambda+{\overline\Lambda})$ 
$(\times 10^{-4})$ & 
$20.2\pm1.0$ & $7.1\pm0.7$ &  $13.1\pm1.2$ \\ 

($\Lambda|{\overline\Lambda})/(\Lambda+{\overline\Lambda})$ ($\times 10^{-2}$) & $1.01\pm0.01$ &
$0.67\pm0.01$ & $0.34\pm0.02$ \\

($\Lambda {\overline \Lambda}|)/(\Lambda+{\overline\Lambda})$ 
($\times 10^{-2}$) & 
$1.16\pm0.01$ & $1.17\pm0.01$ & $-0.01\pm0.01$\\ 

($\Lambda {\Lambda}|)/(\Lambda+{\overline\Lambda})$ ($\times 10^{-5}$) & 
$3.93\pm2.10$ & $3.73\pm1.27$ & $0.20\pm2.45$\\ 

($\Lambda|{\Lambda})/(\Lambda+{\overline\Lambda})$ ($\times 10^{-5}$) & 
$33.0\pm5.1$ & 
$41.2\pm4.1$ & $-8.2\pm6.6$ \\ \hline

Maximum ($\Lambda|{\overline\Lambda}$) (np) background & $4820\pm82$ & 
$6519\pm91$ & \\

Tagged MC ($\Lambda|{\overline\Lambda}$) from 
 ($\Lambda_c|{\overline\Lambda_c}$) evts. &  & 
$477\pm22$ & \\

($\Lambda|{\overline\Lambda}$) (np) estimate, corrected & 
$4466\pm84$ & $6042\pm93$ & \\

Maximum 
($\Lambda|{\overline\Lambda}$) (c\=c) background & 
1671$\pm$221 & \\ 
 
{\bf ($\Lambda|{\overline\Lambda}$) (q\=q), max. bkgnds.}
& 872$\pm$288 & & \\ \hline

($\Lambda|{\overline\Lambda_c}$) evts.
from $({\overline D}\Lambda)|\Lambda_c$ &
$200\pm34$ & $169\pm13$ (tagged) & \\

($\Lambda|{\overline\Lambda}$) (c\=c), corrected
for ($(D\Lambda)|{\Lambda_c}$)  & 1251$\pm$223 & &  \\

{\bf $(\Lambda|{\overline\Lambda})$ (q\=q), ($\Lambda|{\overline\Lambda}$)
(c\=c) 
corrected} & 1290$\pm$371 & & \\ \hline

{\bf
($\Lambda|{\overline\Lambda}$) (q\=q), ALL corrections} & $1643\pm372$ & 
 & \\ \hline
\end{tabular}
\end{small}
\end{center}
\end{table}

Comparing the Monte Carlo vs. data yields in Table I,
we note:
\begin{enumerate}
\item There is an enhancement, in data, of the number of
$\Lambda_c|{\overline\Lambda_c}$ per event, relative to Monte Carlo
(top line of Table I).
This correlated production was the subject of our previous 
paper\cite{lclcbcorrelations}, and was interpreted as evidence
for correlated production of 
charmed baryons from primary quarks.\footnote{We have loosened
the $\Lambda_c$ selection requirements for this analysis relative to
our previous analysis. This results in approximately 20\%(44\%) larger 
single-tag(double-tag) $\Lambda_c$ reconstruction efficiency.}

\item There is an enhancement in the $\Lambda_c|{\overline\Lambda}$ yield
in data relative to Monte Carlo.
This can largely be attributed to the aforementioned
correlated $\Lambda_c|{\overline\Lambda_c}$ production, in which the
decay ${\overline\Lambda_c}\to{\overline\Lambda}$ results in a
${\overline\Lambda}$ in the hemisphere opposite the 
${\Lambda_c}$. Such events will also result from cases in which
charm is compensated by a meson rather than a baryon:
$\Lambda_cK|{\overline\Lambda}{\bar D}$. 

\item In four-baryon events (events with either $\Lambda\Lambda|$ or
$\Lambda|\Lambda$), both data and Monte Carlo show a preference for
$\Lambda|\Lambda$ over $\Lambda\Lambda|$. This is consistent with a
model in which two $\Lambda{\overline\Lambda}|$ pairs, each with 
$\cos\theta_{\Lambda{\overline\Lambda}}\to$1, are created such that
one $\Lambda{\overline\Lambda}|$ pair is directly opposite the second 
$\Lambda{\overline\Lambda}|$ pair.
The actual source of these events,
although interesting, is
not the focus of our current effort and its
discussion will be deferred until further study.
For now, we note that 
to the extent that
$\Lambda{\overline\Lambda}\Lambda{\overline\Lambda}$
events are contributing to both $\Lambda\Lambda$ and
$\Lambda{\overline\Lambda}$ samples, 
statistical consistency between Monte Carlo 
and data in 
the relative ratios of $(\Lambda\Lambda|)/\Lambda$
and $(\Lambda|\Lambda)/\Lambda$ indicates that
Monte Carlo simulations model such events reasonably well.
\item The number of
 same-hemisphere $\Lambda{\overline\Lambda}|$ correlations,
 divided by the total number of single-tag $\Lambda$ particles, gives
 ratios, respectively for data and for Monte Carlo:
 \[
 {\rm DATA:\ }\frac{(\Lambda\overline\Lambda|)}{(\Lambda+\overline\Lambda)}=
 (1.16\pm0.01)\%
 \]
 \[
 {\rm MONTE\ CARLO:\ }
 \frac{(\Lambda\overline\Lambda|)}{(\Lambda+\overline\Lambda)}
 =(1.17\pm0.01)\%
 \]
 which are in excellent agreement. This agreement gives us confidence
 that the Monte Carlo simulation provides an adequate model of the
 non-primary component, which is expected to dominate the small opening
 angle sample.\footnote{We do not include among the non-primary hadrons
 the feed-down $\Lambda$ decay products of $\Lambda_c$, since they include
 part of the primary hadron.}.
\item By contrast, the opposite hemisphere yields:
 \[
 {\rm DATA:\ }\frac{(\Lambda|\overline\Lambda)}{(\Lambda+\overline\Lambda)}=
 (1.01\pm0.01)\%
 \]
 \[{\rm MONTE\ CARLO:\ }
 \frac{(\Lambda|\overline\Lambda)}{(\Lambda+\overline\Lambda)}
 =(0.67\pm0.01)\%
 \]
  indicate large opening-angle $\Lambda|{\overline\Lambda}$ production in
data at a rate 50\% greater than the Monte Carlo simulation.\footnote{It
was found that setting the JETSET parameter 
$\rho$=0.6, as suggested by OPAL, resulted in a
larger $\Lambda|{\overline\Lambda}$ yield, but also produced
a substantial 
deficit in the same-hemisphere Monte Carlo
$\Lambda{\overline\Lambda}|$ yield compared to data, and thus gave a
considerably inferior match to the $\Lambda{\overline\Lambda}$ fragmentation
component compared to $\rho=0.5$.}
 The total number of observed $\Lambda|{\overline\Lambda}$ events should
 arise primarily from three sources:
 \begin{description}
 \item[a)] our signal of interest: direct primary production in
 $e^+e^-\to q{\overline q}$ ($q=u,\ d$, or $s$) events by a mechanism
 similar to that which leads to the observed
 $\Lambda_c|{\overline\Lambda_c}$ enhancement (designated
 ``($\Lambda|{\overline\Lambda}$)(q\=q)'');
 
 \item[b)] $\Lambda|{\overline\Lambda}$ events due to feeddown from $e^+e^-\to
 c{\overline c}\to\Lambda_c{\overline\Lambda_c}$, with:
 $\Lambda_c\to(\Lambda |{\overline\Lambda_c}\to{\overline\Lambda})$
 (designated ``$\Lambda|{\overline\Lambda}$(c\=c)'') and
 
 \item[c)] non-primary $\Lambda$ baryons 
which do not contain a primary light quark
	and are not $\Lambda_c$ decay products
 (designated ``$\Lambda|{\overline\Lambda}$(np)'').
 \end{description}
\end{enumerate}

 Generally, we will use ``fragmentation'', or ``non-primary'' to denote
 particles not containing primary quarks and particles 
which are not direct weak decay
 products of charmed hadrons.  In contrast to a) and b), fragmentation
 will lead to events containing both $\Lambda$ and ${\overline\Lambda}$
 in the same-hemisphere, as well as opposite-hemispheres.  We now outline
 our subtraction of these background components.

\subsection{Subtraction of
non-primary $\Lambda{\overline\Lambda}$ component}
To estimate the non-primary $\Lambda{\overline\Lambda}$
contribution, we rely heavily
on JETSET 7.4 Monte Carlo simulations.
We use the
following procedure
to subtract $\Lambda{\overline\Lambda}$ (np) production
in our
search for an opposite-hemisphere, correlated,
primary $\Lambda|{\overline\Lambda}$ (q\=q)
signal:
\begin{enumerate}
\item We plot the 
$\cos\theta_{\Lambda{\overline\Lambda}}$ distribution in both
data and Monte Carlo simulations, with 
$\theta_{\Lambda{\overline\Lambda}}$ defined as the
opening angle between $\Lambda$ and ${\overline\Lambda}$.
\item We normalize the Monte Carlo 
$\Lambda{\overline\Lambda}$ yield
in the forward hemisphere ($\cos\theta_{\Lambda{\overline\Lambda}}>$0)
to data, 
and subtract the result from data, 
for the full ($\cos\theta_{\Lambda{\overline\Lambda}}$)
angular distribution. 
As noted previously,
the match (absolutely normalized to the total number of
$\Lambda$'s) between data and 
Monte Carlo simulations is satisfactory in the forward
hemisphere ($\cos\theta_{\Lambda{\overline\Lambda}}>0$), however
there is an under-subtraction of events in the back hemisphere.
This subtraction therefore results in an excess
of back-to-back $\Lambda|{\overline\Lambda}$ events in data
relative to Monte Carlo simulations.
\end{enumerate}

\subsection{Subtraction of $\Lambda_c|{\overline\Lambda_c}$
feeddown component
from $\Lambda|{\overline\Lambda}$ yield}

Based on the 
observed number of: a) $\Lambda_c|{\overline\Lambda_c}$,
b) $\Lambda|{\overline\Lambda}$, and 
c) $\Lambda_c|{\overline\Lambda}$ 
in data, we can calculate the total number of 
$\Lambda|{\overline\Lambda}$(c\=c) correlations, assuming 
that any ${\Lambda}$ opposite a
${\overline\Lambda_c}$ is a ${\Lambda_c}$ decay product (we discuss
below two corrections to these assumptions).
This is done by setting
the probability that both charmed baryons in
an $e^+e^-\to c{\overline c}\to\Lambda_c|{\overline\Lambda_c}$ decay
to lambda baryons, relative to just one decaying to
a lambda baryon, equal to the probability that one decays to
a lambda, relative
to the probability that neither decays to lambdas.
Designating $\Lambda_c\to\Lambda$ as the 
inclusive $\Lambda$ yield from
$\Lambda_c$ decay, this probability condition 
can be written as:

\begin{equation}
{N(\Lambda_c\to\Lambda |{\overline\Lambda_c}\to{\overline\Lambda}) \over
N(\Lambda_c\to\Lambda |{\overline\Lambda_c})}
=
{N(\Lambda_c\to\Lambda |{\overline\Lambda_c})\over
N(\Lambda_c|{\overline\Lambda_c})}
\end{equation}

\begin{figure}[htpb]
\begin{picture}(200,250)
\includegraphics{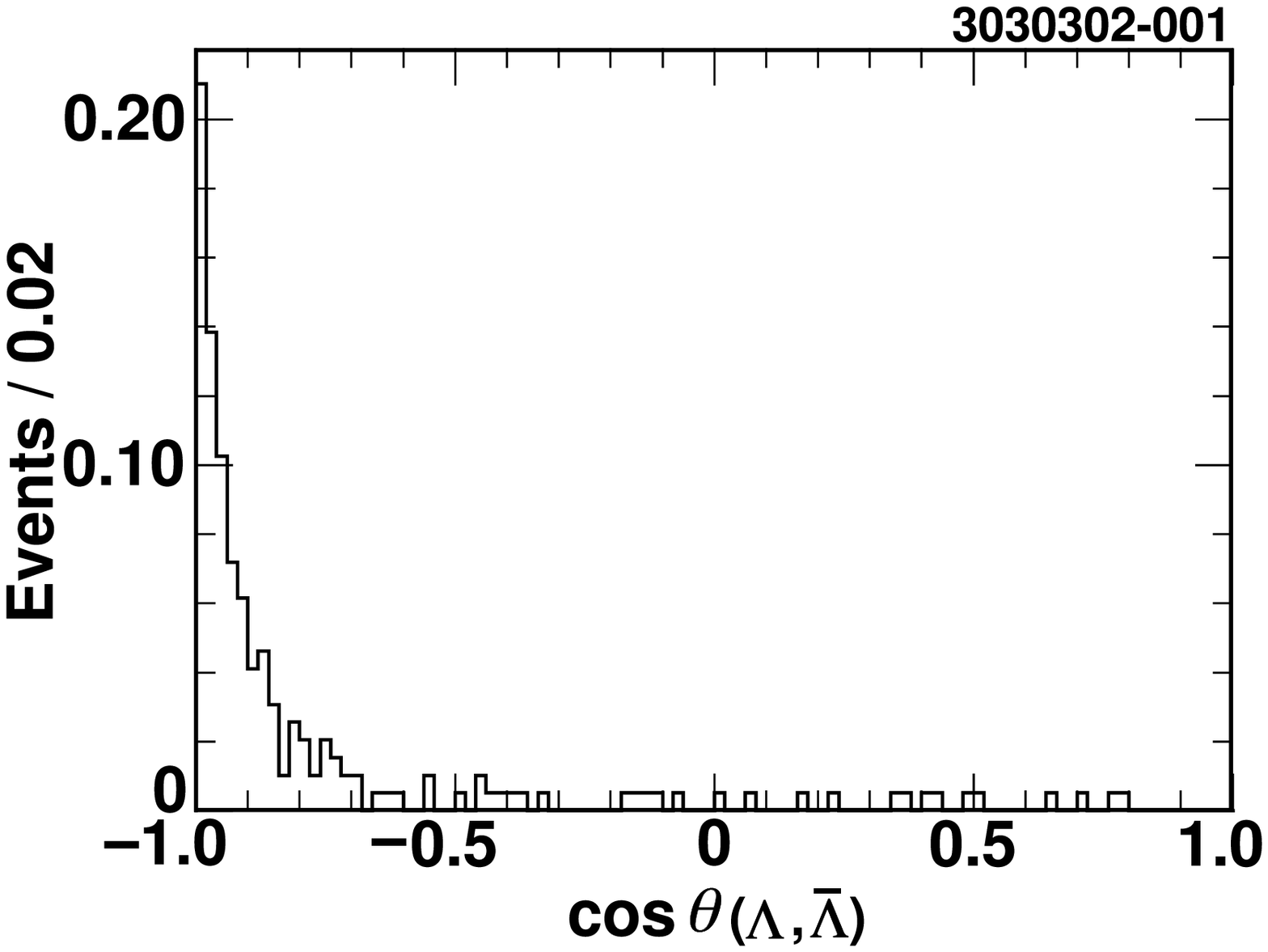}
\end{picture} \\
\caption{\small Cosine of opening angle between
${\overline\Lambda}$ and $\Lambda$ in 
$e^+e^-\to\Lambda_c{\overline\Lambda_c}\to\Lambda {\overline\Lambda}$
events, from Monte Carlo simulations.
The distribution has been normalized to unity. 
\message{put ``MC'' on figure itself.}
}
\label{fig:lclcb_to_l0l0b}
\end{figure}

Monte Carlo simulations indicate that, with the kinematic
requirements we impose on our candidate event sample, $(93\pm2)$\% of
the final state $\Lambda$'s produced in
$\Lambda_c\to\Lambda$ are, indeed, opposite each other.
Figure
\ref{fig:lclcb_to_l0l0b} shows the cosine of the opening angle between the
final state $\Lambda$ relative to the final state ${\overline\Lambda}$ in such
feed-down events; as expected, the distribution peaks at 
$\cos\theta_{\Lambda{\overline\Lambda}}\to -1$.
Thus,
assuming that all the observed ($\Lambda |{\overline\Lambda_c}$)
events result from
$e^+e^-\to\Lambda_c{\overline\Lambda_c}$ 
events in which $\Lambda_c\to\Lambda $ in the
hemisphere opposite the ${\overline\Lambda_c}$, the number
of ``feed-down'' 
($\Lambda_c\to\Lambda |{\overline\Lambda_c}\to{\overline\Lambda }$)
events can be determined from the above equation
directly as:
$N^2(\Lambda |{\overline\Lambda_c})/N(\Lambda_c|{\overline\Lambda_c})$=
$1671\pm221$.\footnote{Note that one must use only one charge conjugate
in the numerator of this ratio ($N(\Lambda|{\overline\Lambda_c})=735.3$) 
to properly compare rates, per reconstructed ${\overline\Lambda_c}$.} 
Using this value for the 
$\Lambda|{\overline\Lambda}$(c\=c) background and subtracting the
scaled MC background as described previously, we obtain, using the yields
from Table I:
 \begin{eqnarray*}
 \lefteqn{N(\Lambda|\overline\Lambda(q\overline q)) =} \\
 & & N(\Lambda|\overline\Lambda)_{data} - N(\Lambda|\overline\Lambda)_{MC}
 \times\frac{N(\Lambda\overline\Lambda|)_{data}}
 {N(\Lambda\overline\Lambda|)_{MC}} -
 N(\Lambda|\overline\Lambda)(c\overline c) =\\
 & & 7362-6519\times\frac{8425}{11395}-1647 = 872\pm288\ {\rm  events}.
 \end{eqnarray*}
 
 Thus, under the most pessimistic of background
 assumptions, we obtain a 3$\sigma$ excess in our estimate of the
 primary, correlated $\Lambda|\overline\Lambda$ yield.   

Some of the observed $\Lambda |{\overline\Lambda_c}$ events
will arise from other sources, such as events in which
a ${\overline\Lambda}$ baryon compensates the baryon number of the $\Lambda_c$
($D{\overline\Lambda}|{\overline K}\Lambda_c$, e.g). There will also
be contributions 
from $(\Xi_c\to\Lambda) |{\overline\Lambda_c}$ and
$(\Omega_c\to\Lambda) |{\overline\Lambda_c}$. The number of
feed-down $(\Lambda_c\to\Lambda |{\overline\Lambda_c}\to{\overline\Lambda})$
calculated through the above prescription 
therefore yields an overestimate of
the true feed-down contribution and therefore will yield a 
lower limit on correlated direct $\Lambda|{\overline\Lambda}$ production
when the feed-down component is subtracted.

We correct our estimate of the
$\Lambda{\overline\Lambda}(c{\overline c})$ background ($1671\pm221$)
with guidance from Monte Carlo simulations.
In Monte Carlo simulations, for which
the parentage of a given particle is known, 
we observe $169\pm13$ 
$\Lambda|{\overline\Lambda}$ events resulting from
$\Lambda_c|{\overline\Lambda}{\overline D}$ events. 
We estimate the number of ($\Lambda|{\overline\Lambda}_c$) events in
our data sample resulting from $\Lambda_c|{\overline\Lambda}{\overline D}$
events (and thus incorrectly attributed to ($\Lambda_c\to\Lambda
|{\overline\Lambda_c}\to{\overline\Lambda}$)) by normalizing the number found
in the Monte Carlo by the ratio of ($\Lambda{\overline\Lambda}_c +
\Lambda_c{\overline\Lambda}$) (all angles) in data relative to 
Monte Carlo simulations:
$$
(\Lambda_c|{\overline\Lambda}{\overline D})_{data} =
\frac{N(\Lambda_c{\overline\Lambda}+\Lambda{\overline\Lambda}_c)_{data}}
{N(\Lambda_c{\overline\Lambda}+\Lambda{\overline\Lambda}_c)_{MC}}\times
N(\Lambda_c|{\overline\Lambda}{\overline D})_{MC} =
\frac{249.4\pm26.2}{211.2\pm28.2}\times 169 = 199.5\pm34.
$$

Using this value, we can now re-calculate the background we attribute to
$\Lambda|{\overline\Lambda}$(c\=c) as: 
$(N(\Lambda |{\overline\Lambda_c}) - 199.5)^2/N(\Lambda_c|{\overline\Lambda_c})=1251\pm223$ events. This smaller background estimate implies
a larger correlated signal yield: 
\[
\Lambda|{\overline\Lambda}(q{\overline q}) = 
7362-[6519\times(8425/11395)]-1251 =
1290\pm371~{\rm events}.
\]
To better estimate the full effect we have to carry out the additional
subtraction described in the following section.

\subsection{Additional correction due to $\Lambda_c|{\overline\Lambda_c}$
production in Monte Carlo simulations.}
Our subtraction of the non-primary component from Monte Carlo
simulations has not been
corrected for known $\Lambda|{\overline\Lambda}$(c\=c) contributions in
the simulation itself.
The opposite-hemisphere $\Lambda|{\overline\Lambda}$ yield from Monte Carlo
simulations given in Table \ref{tab:lamyields} is therefore an overestimate
of the non-primary component that we subtract out. 
Since the parent type in simulations is known, this correction
can be made directly. We tabulate
$477\pm21.8$ $\Lambda{\overline\Lambda}$(c\=c) events
contributing to our $\Lambda|{\overline\Lambda}$ sample in Monte
Carlo simulations. The calculated yield of primary, correlated
$\Lambda|{\overline\Lambda}$ events is now:
\[
\Lambda|{\overline\Lambda}(q{\overline q}) = 
7362-[(6519-477)\times(8425/11395)]-1251 = 1643\pm372.
\]

\subsection{Summary of Subtraction Procedure}
\begin{figure}[htpb]
\begin{picture}(200,250)
\includegraphics{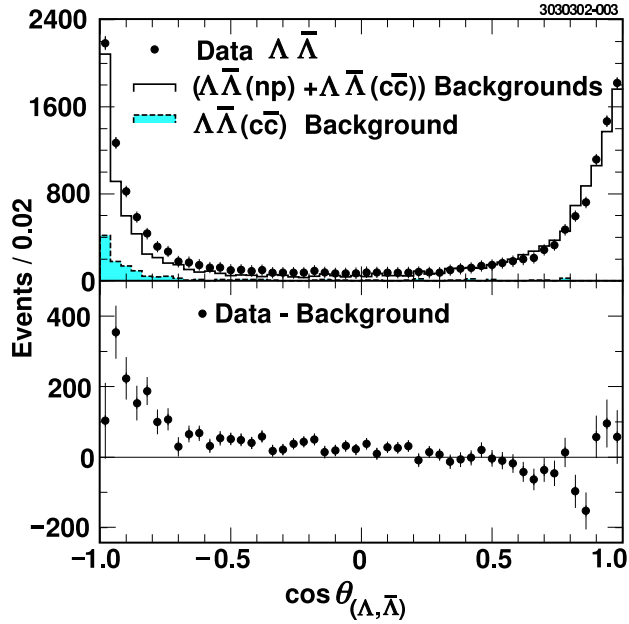}
\end{picture} \\
\caption{\small 
$\cos\theta_{\Lambda{\overline\Lambda}}$ distribution for
data ($\bullet$) 
with total background overlaid (unshaded histogram). The
Monte Carlo $\cos\theta_{\Lambda{\overline\Lambda}}$ distribution
is corrected for the expected
($\Lambda_c\to\Lambda|{\overline\Lambda_c}\to{\overline\Lambda}$)
contribution (normalization obtained from data, with the shape
taken from Monte Carlo simulations), applying all corrections
cited in the text. That total background distribution has
been normalized to the data in the region 
 $\cos\theta_{\Lambda{\overline\Lambda}}>0$, then subtracted from
the data distribution. The bulk of the background histogram is 
due to $\Lambda|{\overline\Lambda}$ (np) production.
The component of the total background due exclusively to
$\Lambda|{\overline\Lambda}$ (c\=c) is shown as the shaded histogram.
The final, background-subtracted (all corrections
applied) data excess is shown in the lower panel.}
\label{fig:costheta_allcuts.ps}
\end{figure}

Figure
\ref{fig:costheta_allcuts.ps} displays the 
$\cos\theta_{\Lambda{\overline\Lambda}}$ distributions, as well as the
subtraction procedure.
Overlaid with the data 
$\cos\theta_{\Lambda{\overline\Lambda}}$ distribution is the 
Monte Carlo simulation.
Applying the maximum possible ${\overline\Lambda}|\Lambda$(c\=c) background
estimate to the data distribution shown in Fig.
\ref{fig:costheta_allcuts.ps}, (i.e.,
not correcting for ${\overline D}\Lambda_c{\overline\Lambda}$ background
in data or ${\overline\Lambda}|\Lambda$(c\=c) in Monte Carlo simulations),
we obtain the most conservative
back-hemisphere excess of $872\pm288$ events (statistical
errors only). Applying all corrections (as in the Figure), 
this excess is approximately doubled.

\subsection{Search for correlated $\Lambda|{\overline\Lambda}$ excess
relative to $\Lambda{\overline K}$}
In our study of $\Lambda_c|{\overline\Lambda_c}$ correlations, we compared
$\Lambda_c|{\overline\Lambda_c}$ to $\Lambda_c|{\overline D}$. In an
analogous way, we shall attempt to compare
$\Lambda|{\overline\Lambda}$ production relative to 
$\Lambda|K$ production. 
To normalize properly,
we have compared the fractional production rate: 
$(\Lambda|{\overline\Lambda})/{\overline\Lambda}$ relative to
$(\Lambda|K)/K$, where the denominator
designates the total number of 
detected single-tag ${\overline\Lambda}$ (or $K$). 
This technique was used to search for evidence of
correlated $\Lambda_c|{\overline\Lambda_c}$ production in our previous
publication. While a ${\overline D}$ tag always contains a primary quark,
our $K$ sample includes non-primary kaons, including
those that are decay products of charmed hadrons.\footnote{There 
are further complications arising
from cases such as: $e^+e^-\to\Lambda{\overline\Lambda}$(1420);
${\overline\Lambda}$(1420)$\to{\overline p}K^+$. In such cases, although
observation of the $\Lambda$ opposite the $K^+$ would be interpreted
as a primary baryon-meson correlation, the true underlying event is a 
primary baryon-antibaryon
correlation. We can safely neglect such instances as long as
the Monte Carlo simulation is producing such events at approximately
the correct rate.}
Although, in principle, the $D|K$ yield can
provide some guidance as to the $c\to K$ rate, there are
still considerable uncertainties resulting from the exact admixture of
non-primary kaons relative to
$c\to K$. What must be true, however, is that 
$s{\overline s}\to\Lambda|K$ events, in which the 
$\Lambda$ and $K$ both contain the primary quarks will result in only
one flavor correlation; non-primary contributions can give rise to both
$\Lambda|K$ and $\Lambda|{\overline K}$
correlations. Even in the absence of correlated primary 
$s{\overline s}\to\Lambda|K$ production,
however, we still expect, because of strangeness conservation, 
$\Lambda$ production in association
with $K$ (and not ${\overline K}$).
For this analysis, we restrict ourselves to continuum
data to exclude possible $B\to K X$ contributions.
Yields are summarized in Table \ref{tab:Kyields}.

\begin{table}
\caption{\label{tab:Kyields}
Kaon-Lambda correlation
yields in data vs. Monte Carlo simulations.
The data are drawn from the same sample as for the previous Table.
Derivation
of final results is discussed in detail in the text. No fake subtractions
have been applied to the quoted charged kaon yields.} 
\begin{center}
\begin{small}
\begin{tabular}{lcc} & Data & Monte Carlo 
\\  \hline
$K^-$ (Continuum, ONLY) & $325427\pm570$ & $1130987\pm1064$ 
\\ 
$K^+$ (Continuum, ONLY) & $330010\pm574$ & $1149240\pm1072$ 
\\ \hline\hline 

($\Lambda|K^+$) & $2009.5\pm50.6$ & $9426.3\pm106.0$ \\
($\Lambda K^+|$) & $738.8\pm30.5$ & $2598.4\pm55.2$ \\

($\Lambda|K^-$) & $979.6\pm38.7$ & $3933.6\pm75.4$ \\
($\Lambda K^-|$) & $344.8\pm23.2$ & $899.4\pm37.6$ \\ \hline \hline

$(\Lambda|K^+)/K^+$ ($\times 10^{-3}$) & $6.1\pm0.1$ & $8.2\pm0.1$ 
\\ 
$(\Lambda K^+|)/K^+$ ($\times 10^{-3}$) & $2.2\pm0.1$ & $2.3\pm0.1$ 
\\ 
$(\Lambda|K^-)/K^-$ ($\times 10^{-3}$) & $3.0\pm0.1$ & $3.5\pm0.1$ 
\\
$(\Lambda K^-|)/K^-$ ($\times 10^{-3}$) & $1.1\pm0.1$ & $0.8\pm0.1$ 
\\ \hline \hline
${\cal D}_{K^+}$ $(\times 10^{-3})$ & $3.9\pm0.2$ & $5.9\pm0.1$ \\ \hline
${\cal D}_{K^-}$ $(\times 10^{-3})$ & $2.0\pm0.1$ & $2.7\pm0.1$ \\ \hline
${\cal S}$ & $8.8\pm0.3$ & $3.8\pm0.1$ 
\\ \hline \hline
\end{tabular}
\end{small}
\end{center}
\end{table}

We have searched for evidence of correlated production of 
$(\Lambda|{\overline\Lambda})$ by comparing the ratio of
$(\Lambda|{\overline\Lambda})/{\overline\Lambda}$ to
$(\Lambda|K)/K$; in each case, we can compare with ``wrong-flavor''
combinations
($(\Lambda|{\Lambda})/{\Lambda}$ and $(\Lambda|{\overline K})/{\overline K}$),
as well as same hemisphere combinations,
as estimates of the fake-kaon
and non-primary components. 
Contributions to 
$\Lambda K^+$ will arise from: i) correlated, primary production, 
ii) weak decays of charmed hadrons in $\Lambda_c|{\overline D}$ events,
iii) compensation of non-primary strange quark production, and 
iv) fake kaons.
Contributions to $\Lambda K^-$ will arise primarily
from non-primary and 
fake kaons. In order to isolate correlated primary production, we
therefore define, for both Monte Carlo simulations as well as data,
the subtracted, normalized fractions:
${\cal D}_{K^+} \equiv (\Lambda|K^+)/K^+ - (\Lambda K^+|)/K^+$ and
${\cal D}_{K^-}\equiv (\Lambda|K^-)/K^- - (\Lambda K^-|)/K^-$. We also define a
signal ratio ${\cal S}$, in analogy to our previous publication
${\cal S}\equiv{({\Lambda|{\overline\Lambda}-
\Lambda|{\overline\Lambda}(c{\overline c}))/\Lambda}\over{
{\cal D}_{K^+}-{\cal D}_{K^-}}}$.
Note that, in constructing this ratio, we subtract the contributions
to the $\Lambda|{\overline\Lambda}$ enhancement from 
$\Lambda_c\to\Lambda|{\overline\Lambda_c}\to{\overline\Lambda}$ to 
ensure that any observed enhancement is not due to the previously
measured $\Lambda_c|{\overline\Lambda_c}$ enhancement.
We find ${\cal S}(data)=8.77\pm0.26$, compared with ${\cal S}(MC)=3.79\pm0.08$
(statistical errors only). This is consistent with an enhancement of
correlated primary baryon-antibaryon production compared to correlated
primary baryon-meson production. We note that the (unevaluated)
systematic uncertainties in
this analysis are expected to be considerable, as indicated by the
discrepancy in the ${\cal D}_{K^+}$ values between data and Monte Carlo
simulations (suggesting a greater likelihood of strangeness conservation
to occur by production of mesons vs. baryons in simulations), as well
as the discrepancy in the ${\cal D}_{K^-}$ values (suggesting different
fake rates in the simulation compared to the data).
Because of the large systematic errors
associated with this exercise, the significance of the
difference between ${\cal S}$ (data)
and ${\cal S}$ (Monte Carlo) cannot be simply evaluated on the basis
of the statistical errors. 
This exercise is to be viewed 
only as a check of the primary $\Lambda|{\overline\Lambda}$ production 
enhancement discussed in
subsections III A through III E.

\subsection{Comparison to correlated $\Lambda_c|{\overline\Lambda_c}$
production}
In principle, one might hope to compare the yield of correlated
$\Lambda|{\overline\Lambda}$ production to that of
 $\Lambda_c|{\overline\Lambda_c}$. Quantitatively, one could compare
correlated primary baryon
production for charmed vs. non-charmed baryons, via the ratio: 
N($\Lambda_c|{\overline\Lambda_c}$)/$\epsilon_{\Lambda_c}^2$/c\=c relative to:
N($\Lambda|{\overline\Lambda}$)/$\epsilon_\Lambda^2$/q\=q. The 
efficiency factors $\epsilon$ include both the reconstruction efficiency
in each case, as well as the fraction of the momentum spectrum accepted,
given the minimum momentum requirements in each case ($p>2.3$ GeV/c for the
$\Lambda_c$ analysis vs. $p>1.0$ GeV/c in the $\Lambda$ analysis).\footnote{
One might
argue that since, in the scaled variable $\beta=p/m$, the two 
requirements are approximately equivalent, the
acceptances are also comparable.} What is unknown in the case of the
$\Lambda$ is what fraction of u\=u vs. d\=d vs. s\=s evolve into
p$|$\=p, n$|$\=n, $\Lambda|{\overline\Lambda}$, $\Lambda|{\overline\Sigma^+}$,
$\Lambda|K^+{\overline p}$, etc. Nevertheless, ignoring such unknowns, and
using efficiencies $\epsilon_\Lambda$ and $\epsilon_{\Lambda_c}$ from Monte
Carlo simulations ($\epsilon_\Lambda\sim0.091$ and 
$\epsilon_{\Lambda_c}\sim0.022)$, we expect to observe $\sim$5300 
$\Lambda|{\overline\Lambda}$ correlations, scaling from the observed
$\Lambda_c|{\overline\Lambda_c}$ correlations. We observe 
$\approx$20\% of the expected value, consistent with a model in which 
primary baryon-antibaryon correlations in s\=s events are populated equally by
$\Lambda|{\overline\Sigma}$, 
$\Lambda|{\overline\Lambda}$ and $\Sigma|{\overline\Sigma}$,
with very little contribution from $e^+e^-\to$u\=u and $e^+e^-\to$d\=d.
Unfortunately, a) photon-finding ($\Sigma\to\Lambda\gamma$
or $\Sigma^+\to p\pi^0$) systematics, 
b) substantially lower signal-to-noise ratios compared
to $\Lambda\to p\pi^-$, c) reduced statistics, 
and d) the necessity to reconstruct a secondary vertex
from a single charged track plus neutrals ($\Sigma^+\to p\pi^0$,
e.g.) make the
$\Sigma$ correlation analyses considerably more difficult than the analysis
described herein.

\section{Systematics}
\begin{figure}[htpb]
\begin{picture}(200,250)
\includegraphics{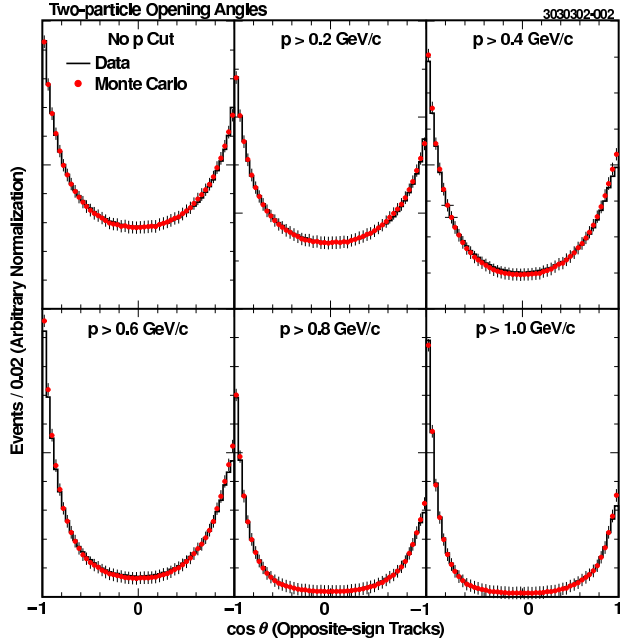}
\end{picture} \\
\caption{\small $\cos(\theta)_{+-}$, defined as the opening
angle distribution for all oppositely-signed charged track pairs, 
for data (+) vs. JETSET 7.4 Monte Carlo simulations (-). The histograms
have been normalized to equal areas.}
\label{fig:cont_v12}
\end{figure}

Since this analysis relies crucially on the ability of the Monte Carlo
simulation
to model angular correlations in the data, it is important that related
kinematic parameters be checked.
To verify that JETSET models two-particle opening angle
($\cos\theta_{+-}$) distributions adequately, we have compared  
Monte Carlo expectations to the data 
for the opening angle distribution between 
oppositely signed tracks (Figure \ref{fig:cont_v12}).
For six different minimum momentum requirements, the Monte Carlo
simulation is
observed to model the data quite well. 

To examine the dependence of our result on the minimum momentum
requirement, we have analyzed the 
$\Lambda|{\overline\Lambda}$
excess (data - background) using different minimum momentum restrictions
on our $\Lambda$ sample. We find that
the match between data and simulation, for the normalized
same-hemisphere yield ($\Lambda{\overline\Lambda}|)/\Lambda$, remains
excellent for $p>0.5$ GeV/c ($0.0164\pm0.0001$ for data
vs. $0.0168\pm0.0001$ for simulation) as well as
for $p>1.2$ GeV/c ($0.0092\pm0.0001$ for data vs.
$0.0091\pm0.0001$ for simulation).
With all corrections applied, opposite-hemisphere
excesses are still observed for
$p_\Lambda>$0.5 GeV/c ($3306\pm543$ events) and 
$p_\Lambda>$1.2 GeV/c ($865\pm283$ events).

Although the minimum momentum requirement ($p_\Lambda>$1.0 GeV/c) should
be highly efficient at removing backgrounds from $B\to\Lambda$X, 
there is still some residual contamination of our single-tag
$\Lambda$ and our double-tag $\Lambda{\overline\Lambda}$ sample from
$\Upsilon$(4S) decays. Inspection of Table I indicates that these
may represent $\sim$3\% corrections to both the single-tag and double-tag
$\Lambda|{\overline\Lambda}$ samples; our yields indicate that the
contamination to the same-hemisphere $\Lambda{\overline\Lambda}|$ yield
is considerably less. Therefore, making such corrections explicit would
have the effect of modifying our ratios and our corrected yields by, at
most, $\sim$3\%.

The analysis would obviously be cleaner if the parent of each
$\Lambda$ could be unambiguously identified.
Because it is not possible to definitively distinguish
$\Lambda$ baryons which contain primary quarks from non-primary baryons,
the extraction of the correlated, primary
$\Lambda|{\overline\Lambda}$ signal is inherently Monte Carlo 
dependent. Only in the previous case of 
$\Lambda_c|{\overline\Lambda_c}$ production
could one conclusively distinguish first-rank from higher-rank baryon
production (since higher-rank $\Lambda_c$ production, at our energies,
is expected to be zero).
In principle, one might hope to separate primary
$\Lambda$ production from non-primary $\Lambda$ production through
several techniques; each technique is, however, fraught with its
own particular difficulties. Measurements such as three-fold
${\overline\Lambda}{\overline D}\Lambda_c$ correlations or 
$e^+e^-\to\gamma\Lambda{\overline\Lambda}X$ production, in which the
(initial state radiation)
photon has sufficiently high energy to exclude $\Lambda_c\to\Lambda X$ 
production, can help us assess, e.g., the contribution to $\Lambda$ production
from charmed baryon decays, but cannot define the presence or
absence of a correlated primary signal.
In the CLEO-c era ($\sqrt{s}\sim$4 GeV), the limited phase
space should make this measurement considerably simpler.

\section{$\Xi|{\overline\Xi}$ correlations}
\begin{figure}[htpb]
\begin{picture}(200,250)
\includegraphics{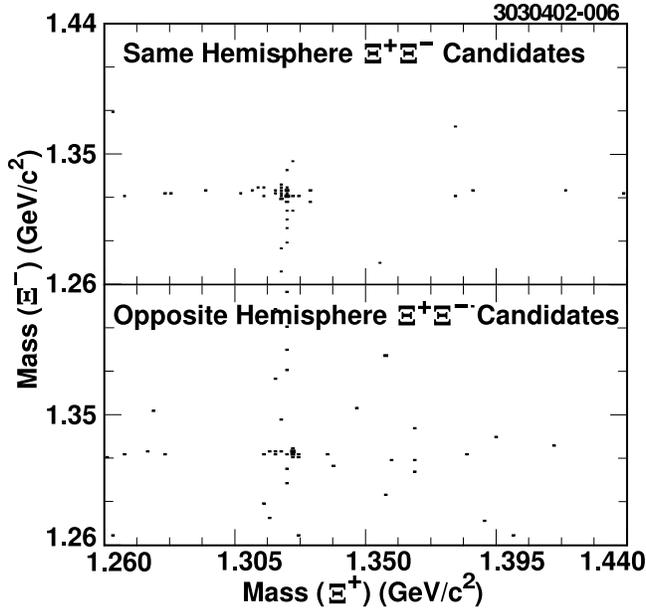}
\end{picture} \\
\caption{\small Same hemisphere and opposite hemisphere $\Xi^+\Xi^-$
candidate invariant mass distributions ($p_{\Xi}>$0.5 GeV/c). 
The correlation is most pronounced
in the same hemisphere.}
\label{fig:xi+xi-}
\end{figure}

In principle, ($\Xi|{\overline\Xi}$) pairs may be used to
further refine our understanding of primary quark-antiquark production and
may, e.g., be used as a discriminant between 
different\cite{popmodels} models
of baryon-antibaryon correlations.
Examining $\Lambda\pi^-$ invariant mass combinations,
we have observed a small, but statistically significant signal for
$\Xi|{\overline\Xi}$ correlations (Figure \ref{fig:xi+xi-}). 
Unfortunately,
the limited signal size is insufficient to attempt to measure primary
correlated $\Xi|{\overline\Xi}$ production.
Nevertheless, it is of interest to compare the yield of
$\Xi^+\Xi^-$, normalized to the total number of charged 
cascades: $\Xi^+\Xi^-/(\Xi^+ + \Xi^-)$, relative to the 
corresponding value for lambda baryons: 
$\Lambda{\overline\Lambda}/(\Lambda + {\overline\Lambda})$.
Loosening our minimum momentum requirement to
0.5 GeV/c, we find $14537\pm135$ total ($\Xi^++\Xi^-)$,
an opposite-hemisphere yield
$(\Xi^+|\Xi^-)=13.0\pm3.9$, and a same hemisphere yield 
$(\Xi^+\Xi^-|)=21.2\pm5.1$. Correspondingly, we obtain
$(\Xi^+\Xi^-|)/(\Xi^+ + \Xi^-)=(1.5\pm0.4)\times 10^{-3}$ for
same-hemisphere cascades, vs. 
$(\Lambda{\overline\Lambda})|/(\Lambda + 
{\overline\Lambda})=(23\pm0.4)\times 10^{-3}$ for same-hemisphere
lambdas. For opposite hemisphere baryon-antibaryon 
correlations, the corresponding
numbers are $(0.9\pm0.3)\times 10^{-3}$ 
and $(28\pm0.6)\times 10^{-3}$. In contrast to
di-lambda production, di-cascade production favors (albeit with small
statistics) same- rather than opposite-hemisphere production.
Normalized to the
total number of baryons, the di-cascade rate (integrated over all angles)
is apparently suppressed relative to
the di-lambda rate. 
This is consistent with a model in which cascade and
lambda production is dominated by light quark popping;
in such a picture,
$\Xi^-$ suppression is therefore a direct consequence of strangeness
suppression.

\section{Summary}
Under conservative assumptions, we observe a 
$\sim 3\sigma$ $(872\pm288)$
excess of opposite-hemisphere $\Lambda|{\overline\Lambda}$ production in
data compared to 
the expectations
of the JETSET 7.4 event
generator combined with the full simulation of our detector, and
after accounting
for feeddown production of 
$\Lambda{\overline\Lambda}$ from
charmed baryons.
With appropriate 
corrections applied, this excess increases to $(1643\pm372)$ events.
These results are consistent with enhanced correlated, primary
$\Lambda|{\overline\Lambda}$ production, of the type observed previously
in $\Lambda_c|{\overline\Lambda_c}$ correlations. 
However, we stress the
inherent Monte Carlo dependence of this conclusion (not present in
the $\Lambda_c|{\overline\Lambda_c}$ correlation 
analysis\cite{lclcbcorrelations}), and that the complete parameter
space of the event generator has not been fully explored. 
Data-taking planned for CLEO-c at $D|{\overline D}$ threshold should
be able to more definitively measure such $\Lambda|{\overline\Lambda}$
correlations, in a considerably less Monte Carlo-dependent manner.

\section{Acknowledgments}

We gratefully acknowledge the effort of the CESR staff in providing us with
excellent luminosity and running conditions.
M. Selen thanks the PFF program of the NSF and the Research Corporation, 
and A.H. Mahmood thanks the Texas Advanced Research Program.
This work was supported by the National Science Foundation, and the
U.S. Department of Energy.

\end{document}